\documentclass[conference]{IEEEtran}

\usepackage{cite}
\usepackage{amsmath,amssymb,amsfonts}
\usepackage{algorithmic}
\usepackage{graphicx}
\usepackage{textcomp}
\usepackage{xcolor}
\usepackage{siunitx}
\def\BibTeX{{\rm B\kern-.05em{\sc i\kern-.025em b}\kern-.08em
    T\kern-.1667em\lower.7ex\hbox{E}\kern-.125emX}}
\begin{document}

\title{Fragment-Level Macro-Diversity Reception in LoRaWAN Networks with LR-FHSS}

\author{\IEEEauthorblockN{1\textsuperscript{st} Samer Lahoud}
\IEEEauthorblockA{\textit{Faculty of Computer Science} \\
\textit{Dalhousie University}, Halifax, Canada\\
sml@dal.ca
}
\and
\IEEEauthorblockN{2\textsuperscript{nd} Kinda Khawam}
\IEEEauthorblockA{\textit{Laboratoire DAVID} \\
\textit{Université de Versailles }, Versailles, France\\
kinda.khawam@uvsq.fr
}

}
\maketitle

\begin{abstract}
The rapid expansion of Internet of Things (IoT) deployments demands wireless protocols that combine high scalability with robust performance. Long Range–Frequency Hopping Spread Spectrum (LR-FHSS) extends LoRaWAN by increasing capacity and resilience through frequency hopping and redundancy. However, current deployments require packet reconstruction at a single gateway, limiting the benefits of LR-FHSS. This paper proposes a macro-diversity reception strategy where multiple gateways collectively receive and combine payload fragments. We develop a stochastic geometry-based analytical model that captures the impact of header repetition, payload fragmentation, and coding redundancy. Closed-form expressions quantify success probabilities under interference, and numerical evaluations demonstrate significant capacity gains over nearest-gateway reception. These results highlight the potential of fragment-level macro-diversity to improve scalability and reliability in future LPWAN deployments.
\end{abstract}


\section{Introduction}
The Internet of Things (IoT) continues to expand rapidly, with applications ranging from smart agriculture and industrial monitoring to infrastructure management and environmental sensing. Supporting this growth requires communication protocols that can sustain long-range connectivity, tolerate interference, and operate at minimal energy cost. Low Power Wide Area Networks (LPWANs), and in particular LoRaWAN, have become foundational to these deployments due to their low power operation and extended coverage in unlicensed sub-GHz bands. However, as IoT networks scale to thousands or millions of devices, LoRaWAN faces increasing performance bottlenecks. The original Chirp Spread Spectrum (CSS)-based LoRa modulation suffers from limited spectral efficiency and high collision rates under dense deployments, largely due to its reliance on pure ALOHA channel access and fixed spreading factors. To address these limitations, Long Range–Frequency Hopping Spread Spectrum (LR-FHSS) has been introduced as a physical layer enhancement for LoRaWAN \cite{semtech2023whitepaper}. LR-FHSS enables devices to transmit using intra-packet frequency hopping over narrowband subcarriers, improving resilience to interference and achieving compliance with duty-cycle and dwell-time constraints through spectral spreading.

LR-FHSS introduces two key mechanisms that differentiate it from LoRaWAN networks using LoRa modulation: (i) header repetition, where multiple replicas of the packet header are transmitted to increase the chance of synchronization; and (ii) payload fragmentation, where the packet is split into smaller fragments transmitted over different frequencies. Payload decoding relies on erasure coding, enabling successful recovery even if some fragments are lost.

While LR-FHSS shows promising improvements in empirical studies, existing analytical evaluations remain limited in scope. Most models assume that a packet must be fully reconstructed at a single gateway, thereby underutilizing the inherent redundancy of LR-FHSS. Moreover, several critical aspects are inadequately addressed: interference is often simplified without accounting for the hopping structure; device and gateway distributions are commonly modeled using grids instead of stochastic processes; and macro-diversity reception, where multiple gateways collaborate to decode a packet, is never considered. These limitations obscure the LR-FHSS performance bounds,  especially in dense networks.

This work proposes a fragment-level macro-diversity reception strategy for LR-FHSS-enabled LoRaWAN, where multiple gateways jointly collect and forward payload fragments, enabling distributed packet reconstruction. To evaluate this strategy, we develop a stochastic geometry-based analytical framework that models interference, spatial randomness, and fragment redundancy. We derive closed-form expressions for the packet success probability under macro-diversity, accounting for key LR-FHSS parameters such as header repetition, payload fragmentation, and coding rate. We validate our model through numerical evaluations comparing macro-diversity with traditional nearest-gateway reception, across different data rate configurations. The results quantify how macro-diversity improves success probability and goodput, and highlight design trade-offs between redundancy, capacity, and interference tolerance. This analysis lays the theoretical foundation for scalable LPWAN deployments using LR-FHSS.

\section{Related Works}

The introduction of LR-FHSS by Semtech~\cite{semtech2023whitepaper} significantly extends LoRaWAN capabilities by improving spectral efficiency, increasing link budget, and ensuring compliance with regional duty-cycle regulations. Simulation-based evaluations such as~\cite{boquet2021lrfhss} have demonstrated the potential of LR-FHSS to improve network capacity, though under simplified assumptions where collisions are modeled only by time and frequency overlap. More recent works~\cite{maldonado2024, ullah2024exp, sanchez-vital2024energy, bukhari2024understanding, jung2024lrfhss} have examined aspects such as sequence design, transceiver implementations, energy consumption, and behavior under real-world trace analysis. These studies deepen the empirical understanding of LR-FHSS, particularly in direct-to-satellite and terrestrial IoT deployments.

However, most existing evaluations rely on simulation or measurement campaigns, with limited theoretical modeling. Early analytical attempts~\cite{ullah2022, maleki2023} addressed delivery and outage probabilities, but either ignored aggregate interference or approximated it using simplified time-on-air metrics, neglecting fading and spatial randomness.

To rigorously model interference in large-scale deployments, stochastic geometry has become essential. Although widely used in cellular network analysis~\cite{haenggi2012stochastic}, its use in LPWAN and LR-FHSS contexts remains limited. The work in~\cite{hattab2024} modeled distributed reception in a cell-free IoT network, offering early insights into multi-gateway decoding. Similarly, \cite{fraire2023recovering} investigated the decoding of headerless LR-FHSS frames, illustrating the role of fragment-level behavior in reception success.

Macro-diversity reception, in which multiple gateways jointly decode transmissions, has been studied in the context of ALOHA-based LPWANs~\cite{song2019macrodiversity}, showing improved reliability through spatial diversity. However, these studies assume full-packet reception at each gateway and do not account for the specific features of LR-FHSS such as erasure coding, header repetition, and intra-packet frequency hopping. Consequently, the redundancy mechanisms and fragment-level diversity inherent to LR-FHSS remain underexploited in current models.

This work addresses this gap by proposing a macro-diversity reception strategy tailored to LR-FHSS. It enables distributed fragment collection across gateways and supports packet reconstruction via centralized coordination. We develop a stochastic geometry-based framework that captures spatial randomness, interference, and key LR-FHSS design parameters. To the best of our knowledge, this is the first theoretical capacity analysis of fragment-level macro-diversity reception in LR-FHSS-enabled LoRaWAN networks.

\section{LR-FHSS Modulation}

\begin{figure}
	\centering
	\includegraphics[width=0.95\linewidth]{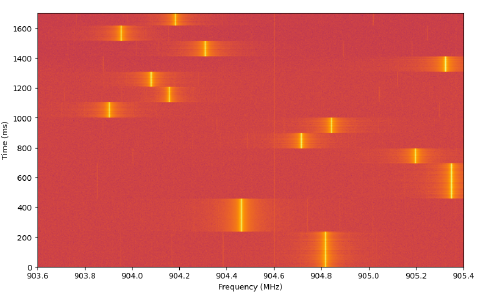}
	\caption{Spectrogram of an LR-FHSS packet showing sequential header replicas followed by pseudo-randomly hopped payload fragments.}
	\label{fig:lr-fhss-spectrogram}
\end{figure}

This section describes the LR-FHSS modulation scheme as deployed in the 902–928 MHz ISM band, which is authorized in Canada and other regions governed by FCC regulations. This configuration is adopted as a representative setting for the analysis. Figure~\ref{fig:lr-fhss-spectrogram} illustrates the time-frequency structure of an LR-FHSS transmission.

\subsection{Data Rates and Coding}

In the FCC band, LR-FHSS supports two standardized data rates: DR5 and DR6. DR5 provides a physical-layer bitrate of 162 bps, uses a convolutional coding rate of \( 1/3 \), and transmits three header replicas. DR6 increases the bitrate to 325 bps with a coding rate of \( 2/3 \) and transmits two header replicas. DR5 offers stronger redundancy for robustness under interference, while DR6 reduces time-on-air and improves spectral efficiency. In both cases, the payload is convolutionally encoded and decoded using a Viterbi decoder to enhance resilience to channel impairments.

\subsection{Packet Structure}

Each LR-FHSS transmission consists of a SyncWord, one or more physical-layer header replicas, and a fragmented payload. The header, transmitted at a fixed rate regardless of the data rate, includes metadata such as the hopping sequence seed, payload length, data rate index, number of header repetitions, and coding rate. Header replicas are transmitted sequentially and enable synchronization with the hopping sequence.

The payload is divided into \( L \) fragments, each assigned to a subcarrier according to a pseudo-random hopping sequence. Successful decoding requires that at least \( \mu L \) fragments be correctly received, where \( \mu \in (0,1) \) is the erasure coding recovery threshold. This fragment-based design enhances reliability without requiring full physical-layer repetition, a key distinction from traditional LoRa modulation.

\subsection{Channelization and Subcarrier Grid}

In the FCC region, each Operating Channel Width (OCW) spans 1.523 MHz and is subdivided into 3,125 Occupied Bandwidths (OBWs), each 488 Hz wide. These OBWs are grouped into 52 frequency-hopping grids. Each grid contains 60 OBWs spaced 25.4 kHz apart, satisfying minimum spectral separation constraints required by FCC regulations.

Each frequency hopping sequence operates entirely within a single grid. This structure ensures spectral spreading and regulatory compliance while maximizing the number of orthogonal hopping patterns available for simultaneous device transmissions.

\subsection{Frequency Hopping Strategy}

Each LR-FHSS device generates a pseudo-random frequency hopping sequence based on a hash of its identifier and fragment index. Hopping is confined to the selected OCW and grid, with successive fragments mapped to different subcarriers in time and frequency. This randomization reduces persistent collisions and mitigates narrowband interference.

To decode payload fragments, a gateway must first decode at least one valid header replica to recover the hopping sequence. In standard deployments, each gateway independently attempts to decode the packet, and successful reception requires that a single gateway collect enough fragments. In contrast, the macro-diversity strategy proposed in this work enables gateways to share recovered headers and collaboratively track and forward fragments, thereby increasing the likelihood of successful packet reconstruction.

\section{System Model}

We consider a LoRaWAN network operating with LR-FHSS at the physical layer. The spatial distributions of gateways and end-devices are modeled as two independent homogeneous Poisson Point Processes (HPPPs), denoted \( \mathcal{P}_g \) and \( \mathcal{P}_d \), with densities \( \lambda_g \) and \( \lambda_d \), respectively. Each device periodically transmits LR-FHSS packets characterized by a fixed number of header replicas \( R \), a number of payload fragments \( L \), a payload coding rate \( \mu \in (0,1) \) that denotes the minimum fraction of correctly received fragments required for successful decoding, and constant transmission power \( P_t \).

\subsection{SINR Model}

Let \( y_k \) denote the distance between a reference device and gateway \( k \). The signal-to-interference-plus-noise ratio (SINR) at gateway \( k \) for message \( m \) (either a header replica or a payload fragment) is expressed as
\begin{equation}
	\mathrm{SINR}_k(m) = \frac{P_t f y_k^{-\alpha}}{N_0 + \sum_{j \in \mathcal{I}(m)} P_t h_j z_{j,k}^{-\alpha}},
\end{equation}
where \( f \) and \( h_j \) are independent Rayleigh fading coefficients (exponentially distributed with unit mean), \( \alpha \) is the path-loss exponent, \( N_0 \) is the thermal noise power, \( \mathcal{I}(m) \) is the set of devices interfering with message \( m \), and \( z_{j,k} \) is the distance between interferer \( j \) and gateway \( k \).

Dividing by \( P_t \), we define normalized noise \( \hat{N}_0 = N_0 / P_t \), leading to the simplified expression
\begin{equation}
	\mathrm{SINR}_k(m) = \frac{f y_k^{-\alpha}}{\hat{N}_0 + \sum_{j \in \mathcal{I}(m)} h_j z_{j,k}^{-\alpha}}.
\end{equation}

A message is successfully decoded if its SINR exceeds the corresponding threshold, denoted \( \sigma_H \) for header replicas and \( \sigma_P \) for payload fragments.

\subsection{Proposed Macro-Diversity Reception Strategy}

We introduce a macro-diversity reception strategy that exploits the fragment-level redundancy provided by LR-FHSS. In conventional LoRaWAN deployments, successful decoding requires that a single gateway decode a header and subsequently receive a sufficient number of payload fragments. In contrast, our proposed strategy enables multiple gateways to collectively participate in fragment reception, following centralized coordination based on successfully decoded headers.

Prior studies~\cite{song2019macrodiversity, hattab2024} have investigated macro-diversity in LPWANs, but without considering LR-FHSS-specific features such as intra-packet frequency hopping and erasure-coded payload fragmentation. Our contribution extends the concept of macro-diversity to the fragment level, enabling distributed reception and decoding across the network.

The proposed architecture incorporates a lightweight local controller that coordinates fragment collection. When a gateway successfully decodes a header replica, it forwards the recovered information to the controller. Upon receiving a valid header, the controller reconstructs the corresponding hopping sequence and issues explicit tracking instructions to all gateways, including those that did not decode the header.

From that point, all gateways begin monitoring the specified hopping sequence and forward any received payload fragments to the controller. Fragment reception is independent across gateways and may result in partial, overlapping, or duplicate coverage of the transmitted fragments. The controller performs de-duplication when necessary and attempts packet reconstruction using the combined set of fragments. Decoding is successful if at least \( \mu L \) distinct fragments are correctly received across the network.

\begin{figure}[ht]
	\centering
	\includegraphics[width=0.9\linewidth]{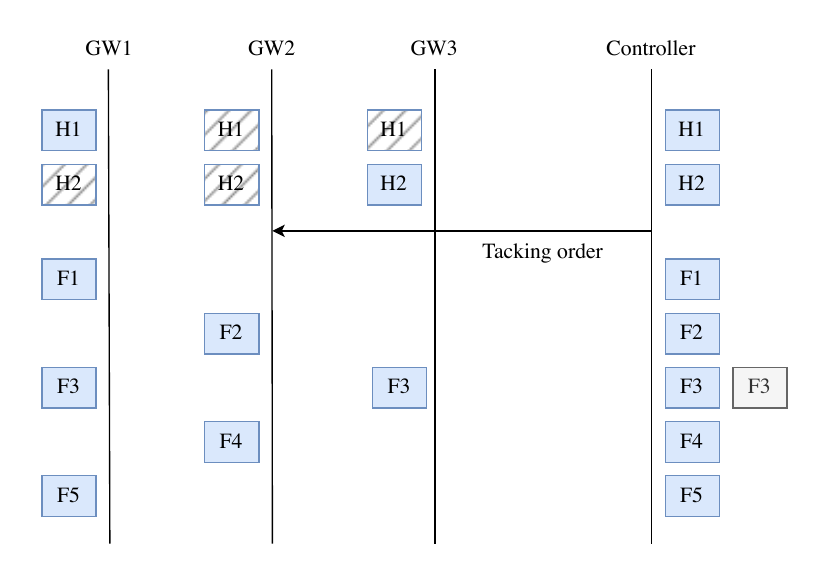}
    \caption{Proposed macro-diversity reception with controller-assisted fragment collection. Gateways that decode a header forward it to the controller, which retrieves the hopping sequence and instructs all gateways to track the corresponding fragments. All received fragments are forwarded to the controller, which performs de-duplication and centralized packet reconstruction.}
	\label{fig:macrodiversity-illustration}
\end{figure}

Figure~\ref{fig:macrodiversity-illustration} illustrates the process. A device transmits a packet with two header replicas (H1 and H2), followed by payload fragments (F1, F2, F3, etc.). Gateways GW1 and GW3 decode different header replicas and forward them to the controller. The controller then extracts the hop sequence and instructs all gateways, including those that did not decode the header, to begin tracking and forwarding fragments. For instance, GW1 receives fragment F1, GW2 receives F2, and both GW1 and GW3 receive F3. The controller de-duplicates fragments when necessary and performs reconstruction using the collected set.

The next section formalizes this strategy through a stochastic geometry-based analytical model and derives tractable performance metrics.

\section{Analytical Model of Macro-Diversity Reception}
\label{sec:analytical-model}
In the proposed macro-diversity architecture, a packet is considered as received successfully if two independent conditions are satisfied:
\begin{enumerate}
	\item At least one gateway correctly decodes one of the \( R \) header replicas transmitted by the device.
	\item At least \( \mu L \) payload fragments are received correctly across all gateways and forwarded to the controller.
\end{enumerate}
These two events are statistically independent, as header decoding enables fragment tracking but does not influence individual fragment receptions once the hopping sequence is known. Thus, the overall packet success probability is
\begin{equation}
	\mathbb{S} = \mathbb{S}^H \cdot \mathbb{S}^P,
\end{equation}
where \( \mathbb{S}^H \) denotes the probability of successful header decoding, and \( \mathbb{S}^P \) denotes the probability of successful payload reconstruction.

\subsection{Header Reception}

Let \( \sigma_H \) denote the SINR decoding threshold for LR-FHSS header replicas, determined by the convolutional coding scheme used at the physical layer. For a typical device transmitting a packet, let \( \mathbb{S}_k^H \) denote the probability that a given gateway \( k \) successfully decodes at least one of the \( R \) header replicas. We define the complementary failure probability as
\begin{equation}
	\bar{\mathbb{S}}_k^H = \Pr\left( \max_{m \in \{1,\ldots,R\}} \mathrm{SINR}_k(m) \leq \sigma_H \right).
\end{equation}
Assuming independent fading and interference across header replicas, and Rayleigh fading with unit mean, we have
\begin{equation}
	\bar{\mathbb{S}}_k^H = \left( \mathbb{E}_{\mathcal{P}_d, h} \left[ 1 - \exp\left( -\sigma_H y_k^\alpha (\hat{N}_0 + I) \right) \right] \right)^R,
\end{equation}
where \( y_k \) is the distance from the device to gateway \( k \), and \( I \) is the aggregate interference from co-channel devices.

In dense LR-FHSS networks, where interference dominates over noise (\( \hat{N}_0 \approx 0 \)), the expression simplifies to
\begin{equation}
	\bar{\mathbb{S}}_k^H = \left( \mathbb{E}_{\mathcal{P}_d, h} \left[ 1 - \exp\left( -\sigma_H y_k^\alpha I \right) \right] \right)^R.
\end{equation}

To model header reception, we consider a thinned Poisson field of interferers. Since transmissions are asynchronous, a header replica is vulnerable to interference from any overlapping transmission in time and frequency. The vulnerability window spans twice the transmission duration, as in pure ALOHA. Frequency overlap is determined by the number of occupied subchannels \( n_{obw} \) across all operating channels \( n_{ocw} \). The effective density of interferers is thus:

\begin{equation}
	\hat{\lambda}_d = \frac{2 \eta_d (R t^H + L t^P)}{n_{ocw} n_{obw}} \lambda_d,
\end{equation}
with \( \eta_d \) the average packet generation rate per device, \( t^H \) and \( t^P \) the durations of a header replica and a payload fragment, and \( \lambda_d \) the nominal device density. The numerator captures the average airtime per device, and the denominator accounts for the spreading of transmissions over the frequency grid.

The Laplace transform of interference evaluated at \( s = \sigma_H y_k^\alpha \) is
\begin{equation}
	\mathcal{L}_I(\sigma_H y_k^\alpha) = \exp\left( -K(\alpha) \hat{\lambda}_d \sigma_H^{2/\alpha} y_k^2 \right),
\end{equation}
where \( K(\alpha) = \frac{2\pi^2}{\alpha \sin(2\pi/\alpha)} \). Substituting, the header decoding probability at gateway \( k \) becomes
\begin{equation}
	\mathbb{S}_k^H = 1 - \left( 1 - \exp\left( -K(\alpha) \hat{\lambda}_d \sigma_H^{2/\alpha} y_k^2 \right) \right)^R,
\end{equation}
which, using the binomial theorem, can be expanded as
\begin{equation}
	\mathbb{S}_k^H = \sum_{r=1}^{R} \binom{R}{r} (-1)^{r+1} \exp\left( -r K(\alpha) \hat{\lambda}_d \sigma_H^{2/\alpha} y_k^2 \right).
\end{equation}

Since the device transmission is broadcast, the header is considered successfully received if at least one gateway decodes one header replica. Thus, the total header success probability is
\begin{equation}
	\mathbb{S}^H = 1 - \mathbb{E}_{\mathcal{P}_g} \left[ \prod_{k \in \mathcal{P}_g} \bar{\mathbb{S}}_k^H \right].
\end{equation}
Applying the probability generating functional of a PPP, and integrating over all possible distances, we get
\begin{equation}
	\mathbb{S}^H = 1 - \exp\left( - \int_0^\infty 2\pi \lambda_g \left(1 - \mathbb{S}_k^H(y)\right) y \, dy \right),
\end{equation}
which simplifies to
\begin{equation}
	\mathbb{S}^H = 1 - \exp\left( \sum_{r=1}^{R} \binom{R}{r} \frac{(-1)^r \pi \lambda_g}{r K(\alpha) \hat{\lambda}_d \sigma_H^{2/\alpha}} \right).
\end{equation}
Defining $K_1(\alpha) = \frac{\pi}{K(\alpha)}, \quad K_2(R) = \sum_{r=1}^{R} \binom{R}{r} \frac{(-1)^r}{r}$, we obtain the closed-form expression
\begin{equation}
	\mathbb{S}^H = 1 - \exp\left( K_1(\alpha) K_2(R) \sigma_H^{-2/\alpha} \frac{\lambda_g}{\hat{\lambda}_d} \right).
	\label{SH}
\end{equation}

\subsection{Payload Reception}

After successful header decoding, the controller issues tracking orders based on the recovered hopping sequence, allowing fragment collection across all gateways. Each payload fragment is transmitted once, without physical-layer repetition. The probability that a given fragment is received correctly is determined by its SINR relative to the decoding threshold \( \sigma_P \), which may differ from \( \sigma_H \). 

Since payload fragments are transmitted independently and experience statistically identical interference and fading conditions as header replicas, we compute the fragment success probability by setting \( R = 1 \) and replacing the decoding threshold \( \sigma_H \) with \( \sigma_P \) in the interference model, yielding
\begin{equation}
	\mathbb{S}_f^P = 1 - \exp\left( -K_1(\alpha) \sigma_P^{-2/\alpha} \frac{\lambda_g}{\hat{\lambda}_d} \right).
\end{equation}

Let \( X \) denote the number of correctly received fragments among the \( L \) transmitted. Since fragment receptions are independent, \( X \) follows a binomial distribution \( X \sim \text{Binomial}(L, \mathbb{S}_f^P) \). Accordingly, successful payload decoding occurs if more than \( \mu L \) fragments are received correctly. Thus, the payload success probability is
\begin{equation}
	\mathbb{S}^P = \Pr(X \geq \lceil \mu L \rceil) = 1 - \sum_{k=0}^{\lceil \mu L \rceil - 1} \binom{L}{k} (\mathbb{S}_f^P)^k (1 - \mathbb{S}_f^P)^{L - k}.
\end{equation}
Substituting the expression for \( \mathbb{S}_f^P \), and defining \[
\gamma = K_1(\alpha) \sigma_P^{-2/\alpha} \frac{\lambda_g}{\hat{\lambda}_d},
\] we obtain
\begin{equation}
	\mathbb{S}^P = 1 - \sum_{k=0}^{\lceil \mu L \rceil - 1} \binom{L}{k} \left( 1 - e^{-\gamma} \right)^k \left( e^{-\gamma} \right)^{L-k}.
    \label{SP}
\end{equation}

Combining the two independent terms in (\ref{SH}) and (\ref{SP}), the final expression for the overall packet success probability is

\begin{equation}
	\mathbb{S} = \mathbb{S}^H \times \mathbb{S}^P.
\end{equation}

\section{Numerical Evaluation}

We evaluate the performance of the proposed macro-diversity reception model in LR-FHSS-enabled LoRaWAN networks and compare it to a baseline \emph{nearest-gateway reception} strategy. In the baseline, only the closest gateway tries to independently decode both the header and a sufficient number of payload fragments. In such case, no cooperation between gateways is allowed.

All performance metrics are computed using the closed-form expressions derived in Section~\ref{sec:analytical-model}. The key simulation parameters are provided in Table~\ref{tab:simulation-parameters}. Two data rates are considered: DR5, which includes strong redundancy via low coding rate and three header replicas, and DR6, which achieves higher spectral efficiency with reduced redundancy. The lower SINR threshold for header decoding reflects the use of convolutional coding with a rate of \( \frac{1}{2} \) applied on each header replica.

The network load is expressed as the \emph{offered load per gateway}, defined as $\eta_d \cdot \lambda_d / \lambda_g \times B_T$, where \( \eta_d \) is the average packet generation rate per device, \( \lambda_d \) and \( \lambda_g \) are the spatial densities of devices and gateways, respectively, and \( B_T \) is the total packet size in bits. This quantity represents the average incoming traffic at a given gateway, in Mbps. 
Intuitively, a low offered load corresponds to scenarios with either sparse device deployments, infrequent transmissions, or dense gateway coverage that distributes the load more evenly. In contrast, a high offered load arises from denser device deployments, higher transmission rates, or sparse gateway infrastructure, all of which increase contention and interference at each gateway.

\begin{table}[ht]
	\caption{Simulation Parameters}
	\label{tab:simulation-parameters}
	\centering
	\begin{tabular}{ll}
		\hline
		\textbf{Parameter} & \textbf{Value} \\
		\hline
		Path loss exponent \( \alpha \) & 3.5 \\
		SINR threshold for header \( \sigma_H \) & \( -22 \) dB \\
		SINR threshold for payload \( \sigma_P \) & \( -20 \) dB \\
		Data rate & DR5, DR6 \\
		Header repetitions \( R \) & 3 (DR5), 2 (DR6) \\
		Payload coding rate \( \mu \) & \( 1/3 \) (DR5), \( 2/3 \) (DR6) \\
        Header size & 114 bits\\
		Payload fragment size & 50 bits \\
		Maximum payload size & 58 bytes (DR5), 133 bytes (DR6) \\
		Channel configuration & 1 OCW, 3120 OBW \\
		\hline
	\end{tabular}
\end{table}

\begin{figure}[ht]
\centering
\includegraphics[width=\linewidth]{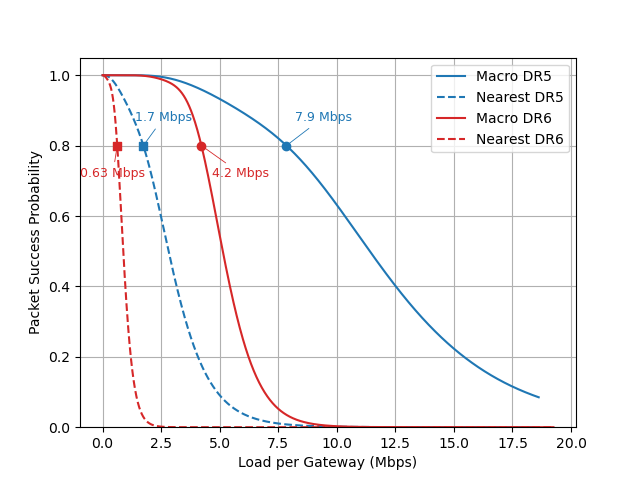}
\caption{Total packet success probability as a function of load per gateway. Macro-diversity significantly extends the reliable operating region for both DR5 and DR6.}
\label{fig:success_total}
\end{figure}

Figure~\ref{fig:success_total} shows the total packet success probability as a function of load per gateway. For DR5, macro-diversity maintains a success probability above 80\% up to 7.9 Mbps, compared to 1.7 Mbps under the nearest-gateway model. For DR6, the threshold under macro-diversity is 4.2 Mbps, while the baseline saturates at 0.63 Mbps. These results show that fragment-level cooperation significantly improves scalability, especially under heavy load.

\begin{figure}[ht]
\centering
\includegraphics[width=\linewidth]{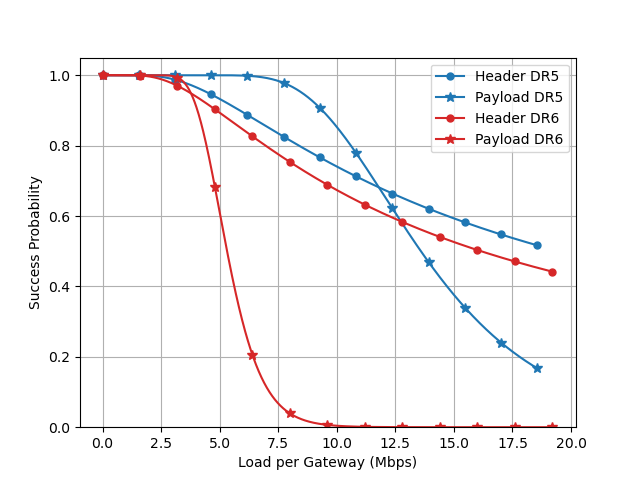}
\caption{Header and payload success probabilities under macro-diversity. At high success probabilities, header decoding is the limiting factor in DR5, while payload decoding limits performance in DR6.}
\label{fig:success_macro}
\end{figure}

To understand the contribution of each decoding stage, Figure~\ref{fig:success_macro} shows the header and payload success probabilities under macro-diversity. For DR5, payload success remains near 100\% up to approximately 7 Mbps. In this range, header decoding is the sole limiting factor, making header repetition critical for reliability. The curves intersect around 12 Mbps, beyond which payload failures begin to contribute significantly to overall degradation.

In contrast, DR6 exhibits earlier payload degradation, starting near 3 Mbps. Payload success drops below 60\% by 5 Mbps, while header success remains higher than 80\% to around 7 Mbps. This creates a region where headers are successfully decoded, but too few fragments are received to reconstruct the payload, rendering header success ineffective.

\begin{figure}[ht]
\centering
\includegraphics[width=\linewidth]{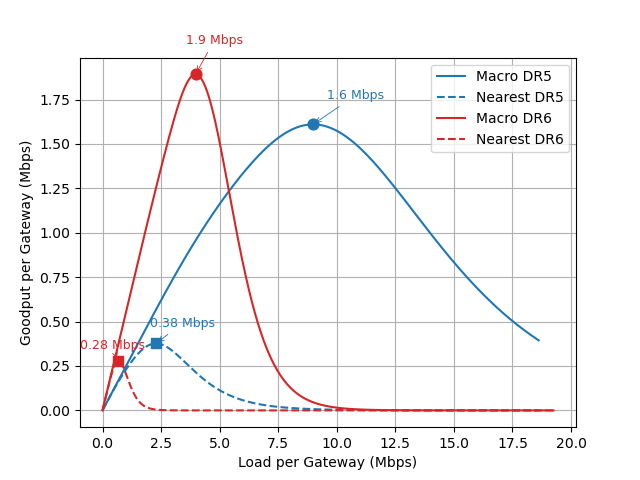}
\caption{Goodput per gateway for macro-diversity and nearest-gateway reception. DR6 peaks higher but saturates earlier; DR5 sustains performance at heavier loads. Macro-diversity improves peak goodput by up to 5 times and extends the sustainable load range.}
\label{fig:goodput}
\end{figure}

Figure~\ref{fig:goodput} compares the \emph{goodput per gateway}, computed as $\mathbb{S} \cdot \eta_d \cdot B_P$, where \( \mathbb{S} \) is the total packet success probability, \( \eta_d \) the per-device packet rate, and \( B_P \) the payload size in bits. Under macro-diversity, DR6 achieves the highest peak goodput at 1.9 Mbps but saturates earlier for increasing load. In contrast, DR5 peaks slightly lower, around 1.6 Mbps, but sustains this level at higher load due to its stronger protection mechanisms. Under the nearest-gateway model, goodput remains below 0.4 Mbps for both data rates, highlighting the severe impact of collisions and lack of cooperation.

These results confirm that fragment-level macro-diversity enhances both the capacity and robustness of LR-FHSS networks. The trade-off between redundancy and efficiency is clearly visible: DR5 offers better protection under high traffic, while DR6 leverages higher spectral efficiency at the cost of sharper performance degradation. These findings highlight the importance of jointly optimizing data rate, redundancy, and cooperation mechanisms when designing scalable LR-FHSS deployments.

\section{Conclusion and Outlook on Macro-Diversity}
This work introduced a fragment-level macro-diversity reception strategy for LR-FHSS-enabled LoRaWAN networks, addressing a key limitation of current deployments that rely on single-gateway packet reconstruction. By allowing multiple gateways to cooperatively forward payload fragments, the proposed strategy fully exploits the redundancy built into LR-FHSS through intra-packet frequency hopping, header repetition, and payload fragmentation. We developed a stochastic geometry model to evaluate header and payload success probabilities under macro-diversity. Closed-form expressions reveal how reliability depends on coding rate, redundancy, and traffic load. Numerical results show that the strategy significantly improves capacity and goodput, while remaining fully compatible with standard LR-FHSS signaling.

However, realizing fragment-level macro-diversity in practice introduces new challenges. Coordinating fragment forwarding requires gateways to operate under asynchronous conditions, manage fragment-to-packet association, and perform de-duplication, all while preserving the energy efficiency and minimal signaling overhead that define LPWANs. These constraints open important directions for future work: designing scalable and standard-compliant aggregation mechanisms, enabling dynamic gateway selection, and developing lightweight control protocols for distributed fragment tracking. Addressing these challenges will be essential to unlocking the full potential of LR-FHSS in large-scale IoT deployments.

\bibliographystyle{IEEEtran}
\bibliography{biblio}
\end{document}